\PassOptionsToPackage{table,xcdraw}{xcolor}
\documentclass[11pt,a4paper]{article}
\pdfoutput=1

\makeatletter
\def\@fpheader{}
\makeatother

\usepackage{jheppub}
\usepackage{gensymb}
\DeclareSymbolFont{matha}{OML}{txmi}{m}{it}
\DeclareMathSymbol{\varv}{\mathord}{matha}{118}
\usepackage{amsmath}
\usepackage{amssymb}
\usepackage{mathrsfs}
\usepackage{graphicx}
\usepackage{subfigure}
\usepackage{pstricks}
\usepackage{bm}
\usepackage{pbox}
\usepackage{placeins}
\usepackage{booktabs}
\usepackage[T1]{fontenc}
\usepackage{footnote}
\usepackage{pdfpages}
\usepackage{hhline}
\usepackage{multirow}
\usepackage{multicol}
\usepackage{enumitem}
\usepackage[toc,page]{appendix}
\allowdisplaybreaks
\usepackage{comment}
\usepackage{float}
\usepackage{slashed}
\usepackage{xcolor}
\usepackage{textcomp}
\usepackage{multirow}
\usepackage[numbers]{natbib}
\usepackage{notoccite}
\usepackage{indentfirst}
\usepackage{jheppub}
\usepackage{gensymb}
\DeclareSymbolFont{matha}{OML}{txmi}{m}{it}
\DeclareMathSymbol{\varv}{\mathord}{matha}{118}
\usepackage{amsmath}
\usepackage{array}
\usepackage{amssymb}
\usepackage{graphicx}
\usepackage{subfigure}
\usepackage{pstricks}
\usepackage{bm}
\usepackage{pbox}
\usepackage{placeins}
\usepackage{booktabs}
\usepackage[T1]{fontenc}
\usepackage{footnote}
\usepackage{pdfpages}
\usepackage{hhline}
\usepackage{multirow}
\usepackage{multicol}
\usepackage{enumitem}
\usepackage[toc,page]{appendix}
\allowdisplaybreaks
\usepackage{comment}
\usepackage{float}
\usepackage{slashed}
\usepackage{xcolor}
\usepackage{textcomp}
\usepackage{multirow}
\usepackage[numbers]{natbib}
\usepackage{notoccite}
\usepackage{soul}
\usepackage{extarrows}
\usepackage{makecell}
\usepackage{longtable}
\usepackage{physics}
\usepackage{comment}
\usepackage{blindtext}
\usepackage{csquotes}
\usepackage{soul}
\usepackage{indentfirst}
\DeclareUnicodeCharacter{0304}{ }
\usepackage{rotating}
\usepackage{tabularx}
\usepackage{extarrows}
\usepackage{physics}
\usepackage{comment}
\usepackage{cancel}
\usepackage{blindtext}
\usepackage{csquotes}
\usepackage[none]{hyphenat}
\DeclareUnicodeCharacter{0304}{ }
\usepackage{rotating}
\usepackage{tabularx}

\usepackage[many]{tcolorbox}
\definecolor{fg}{RGB}{34,139,34}

\renewcommand{\thetable}{\Roman{table}}

\def\figureautorefname~#1\null{Fig.\,#1\null}
\def\equationautorefname~#1\null{Eq.\,(#1)\null}
\def\tableautorefname~#1\null{Tab.\,#1\null}
\definecolor{MyDarkBlue}{rgb}{0.1, 0.1, 0.8} 
\definecolor{MyLightBlue}{rgb}{0.22,0.51,0.9}
\definecolor{MyGreen}{rgb}{0.0, 0.5, 0.0}
\definecolor{BrickRed}{rgb}{0.8, 0.25, 0.33}

\RequirePackage{hyperref}
\hypersetup{colorlinks=true, citecolor=MyGreen,linkcolor=BrickRed, urlcolor=MyLightBlue}
\usepackage{cleveref}
\crefname{equation}{Eq.}{Eqs.} 
\crefrangelabelformat{equation}{(#3#1#4--#5#2#6)}

\title{\bf Higher-Order Corrections to Quantum Observables in $h\to WW^*$}
\author[a]{Dorival Gon\c{c}alves,}
\author[a]{Ajay Kaladharan,}
\author[a]{Alberto Navarro}
\affiliation[a]{Department of Physics, Oklahoma State University, Stillwater, OK 74078, USA}
\emailAdd{dorival@okstate.edu}\emailAdd{kaladharan.ajay@okstate.edu}\emailAdd{alberto.navarro\_serratos@okstate.edu}

\abstract{The Higgs boson decay $h \to WW^* \to \ell^+ \nu_\ell \ell'^- \bar{\nu}_{\ell'}$ provides a unique window into the structure of the Higgs couplings to electroweak gauge bosons and has recently gained attention for its potential to unveil quantum properties such as quantum entanglement between the intermediate gauge bosons. In this work, we present a systematic study of next-to-leading order electroweak corrections to the angular coefficients characterizing this decay. While these coefficients are highly constrained at leading order, radiative corrections induce shifts of up to 5\% to the existing terms and generate novel structures that vanish at leading order, breaking previous relations among coefficients. While higher-order effects influence the results, the two-qutrit quantum structure in the $h\to WW^*$ channel exhibits greater stability under such corrections than in the previously studied $h \to ZZ^*$ decay.}

\begin{document}
\maketitle

\begin{sloppypar}

\section{Introduction}
\label{sec:intro}

Precision studies of Higgs boson decays into electroweak gauge bosons, $h \to ZZ^*,WW^*$,  play a central role in the LHC physics program, offering stringent tests of the Standard Model (SM) and sensitive probes of new physics~\cite{ATLAS:2012yve,CMS:2012qbp,Buszello:2002uu,Godbole:2007cn,Bolognesi:2012mm,Englert:2012xt,Artoisenet:2013puc,Caola:2013yja,Campbell:2013una,Englert:2014aca,Azatov:2014jga,Buschmann:2014sia,Corbett:2015ksa,Goncalves:2017gzy,Brehmer:2017lrt,Goncalves:2017iub,Goncalves:2018pkt,CMS:2019ekd,Gritsan:2020pib,Goncalves:2020vyn,Hall:2022bme,CMS:2022ley,ATLAS:2023dnm,ATLAS:2023mqy,Bhardwaj:2024lyr}. The corresponding angular distributions encode rich information about the nature of the Higgs couplings and potential deviations arising from physics beyond the SM, including novel sources of CP-violation.  Recently, it has been proposed that these angular observables can be interpreted through the lens of quantum information theory~\cite{Barr:2021zcp,Aguilar-Saavedra:2022wam,Ashby-Pickering:2022umy,Aguilar-Saavedra:2022mpg,Aoude:2023hxv,Fabbrichesi:2023cev,Fabbrichesi:2023jep,Fabbri:2023ncz,Bernal:2023ruk,Morales:2023gow,Bi:2023uop,Barr:2024djo,Aguilar-Saavedra:2024whi,Subba:2024mnl,Bernal:2024xhm,Sullivan:2024wzl,Aguilar-Saavedra:2024jkj,Grossi:2024jae,Ruzi:2024cbt,Wu:2024ovc,Ding:2025mzj,DelGratta:2025qyp,Goncalves:2025qem,Ruzi:2025jql}. Within this framework, the reconstruction of the spin density matrix of the diboson system enables the assessment of quantum correlations, most notably entanglement, bringing a compelling new perspective to these studies.\footnote{These quantum observables have also been explored in other LHC processes, most notably in top quark pair production~\cite{Afik:2020onf,Fabbrichesi:2021npl,Severi:2021cnj,Aguilar-Saavedra:2022uye,Afik:2022kwm,Afik:2022dgh,Severi:2022qjy,Aoude:2022imd,Dong:2023xiw,Aguilar-Saavedra:2023hss,Han:2023fci,Sakurai:2023nsc,Cheng:2023qmz,Maltoni:2024tul,Maltoni:2024csn,White:2024nuc,Dong:2024xsg,Dong:2024xsb,Cheng:2024btk,Altomonte:2024upf,Han:2024ugl,Cheng:2025cuv,Afik:2025ejh,Nason:2025hix}, where both ATLAS and CMS have recently reported observation of spin entanglement~\cite{ATLAS:2023fsd,CMS-PAS-TOP-23-001,PhysRevD.110.112016}. Furthermore, the possibility that entanglement plays a fundamental role in the emergence of key symmetries in particle physics has also been recently investigated~\cite{Cervera-Lierta:2017tdt,Beane:2018oxh,Liu:2022grf,Carena:2023vjc,Thaler:2024anb,vonKuk:2025kbv,Carena:2025wyh}.}

In the specific channel $h\to ZZ^*\to e^+e^-\mu^+\mu^-$, higher-order corrections have been shown to induce sizable effects, largely shifting the existing leading order (LO) angular coefficients and introducing novel terms comparable in magnitude to the LO contributions~\cite{Grossi:2024jae,DelGratta:2025qyp,Goncalves:2025qem}. Beyond being crucial for robust new physics searches, these corrections also challenge the two-qutrit interpretation of the system, due to sizable non-factorizable next-to-leading order (NLO) electroweak (EW) contributions~\cite{Goncalves:2025qem}. These findings highlight that higher-order effects are not merely a technical refinement, but a qualitative requirement for any consistent quantum information interpretation of the Higgs decay $h \to e^+e^-\mu^+\mu^-$, whether these corrections are incorporated into the signal modeling or considered part of the background~\cite{Aguilar-Saavedra:2025byk}.

Motivated by these findings, we now turn to the complementary channel $h\to WW^*\to\ell^+\nu_\ell\ell'^-\bar{\nu}_{\ell'}$. This mode is of critical importance both experimentally, owing to its large branching fraction, and theoretically, as its angular observables provide independent sensitivity to anomalous Higgs couplings in an effective field theory framework. Unlike the $Z$ boson, which couples to both left- and right-handed fermions, resulting in a moderate spin-analyzing power, the $W$ boson interacts exclusively with left-handed fermions (and right-handed antifermions). This purely chiral structure leads to maximal spin-analyzing power, making the charged leptons in $W$ decays particularly sensitive probes of the boson’s polarization and, by extension, of the structure of the Higgs-gauge boson interactions. These distinctive features motivate a dedicated assessment of how NLO EW corrections modify both the angular coefficients\footnote{Angular coefficients are crucial observables for precision studies and searches for physics beyond the SM~\cite{ATLAS:2016rnf,ATLAS:2018gqq,Goncalves:2018fvn,Goncalves:2018ptp,ATLAS:2019zrq,CMS:2019nrx,MammenAbraham:2022yxp,Bhardwaj:2023ufl,Denner:2023ehn,Denner:2024tlu,CMS:2024ony,Carrivale:2025mjy,Grossi:2024jae,DelGratta:2025qyp,Goncalves:2025qem}. Hence, it is essential to understand how they are affected by radiative corrections.} and their implications for the validity of the two-qutrit interpretation in the $h \to WW^*$ channel.
 
This paper is organized as follows. In~\autoref{sec:entanglement}, we review the density matrix formalism and introduce concurrence as a measure of bipartite entanglement. In~\autoref{sec:tomography}, we outline the standard quantum tomography framework for reconstructing the $WW$ system at the LHC. In~\autoref{sec:analysis}, we present the leading-order and next-to-leading-order electroweak results for the $h \to WW^* \to \ell^+ \nu_\ell \ell'^- \bar{\nu}_{\ell'}$ process. We derive the angular coefficients, quantify the impact of higher-order effects, and assess their consequences for interpreting the system as a two-qutrit quantum state. A general comparison with the $h \to e^+e^-\mu^+\mu^-$ channel is also provided. Finally, we summarize our findings and provide an outlook in~\autoref{sec:summary}.

\section{Quantum Entanglement}
\label{sec:entanglement}

The density matrix $\rho$ plays a central role in quantum mechanics, encapsulating all observable information about a quantum system. For a general mixed state composed of pure states $\ket{\Psi_i}$ with associated classical probabilities $ p_i$, the density matrix can be written as
\begin{equation}
    \rho = \sum_i p_i \ket{\Psi_i} \bra{\Psi_i}, \quad \text{with} \quad p_i \geq 0, \quad \sum_i p_i = 1.
\end{equation}
A valid physical density matrix must be Hermitian $( \rho^\dagger = \rho )$, normalized $(\text{Tr}(\rho) = 1)$, and positive semi-definite $(\bra{\Psi_i} \rho \ket{\Psi_i} \geq 0 ~\text{for all}~\ket{\Psi_i})$. In particular, these requirements imply that its eigenvalues  $\lambda_i$ satisfy
\begin{equation}
    0 \leq \lambda_i \leq 1, \quad \sum_i \lambda_i = 1.
    \label{eq:eigenvalue}
\end{equation}

Consider now a bipartite system composed of two subsystems, $A$ and $B$. The full system is said to be separable if its density matrix can be written as a convex sum
\begin{equation}
    \rho = \sum_i p_i \rho_A^{i} \otimes \rho_B^{i}, \quad \text{with} \quad p_i \geq 0, \quad \sum_i p_i = 1,
    \label{eq:separable}
\end{equation}
where $\rho_A^{i}$ and $\rho_B^{i}$ are density matrices acting on the Hilbert spaces of $A$ and $B$, respectively. If no such decomposition exists, the system is said to be \emph{entangled}~\cite{PhysRevA.40.4277}.

To quantify entanglement, we use the concurrence measure $\mathcal{C}(\rho)$~\cite{PhysRevA.64.042315,PhysRevLett.80.2245,PhysRevA.54.3824}, which serves as a well-established indicator of non-separability in bipartite quantum systems. In this study, we consider a system composed of two spin-1 particles, the $W$ bosons originating from the Higgs boson decay, $h\to WW^*$, forming a two-qutrit quantum system. For generic mixed states, determining $\mathcal{C}(\rho)$ exactly is a nontrivial task, as closed-form expressions are typically limited to two-level (qubit) subsystems. However, the lower and upper bounds on $\mathcal{C}(\rho)$ can be used to estimate entanglement in higher-dimensional settings. These bounds are defined by~\cite{PhysRevLett.98.140505,PhysRevA.78.042308,Fabbrichesi:2023cev}
\begin{align}
    \left(\mathcal{C}(\rho)\right)^2
    &\geq 
    2\, \max\Big\{ 0,\, 
    \Tr[\rho^2] - \Tr[\rho_A^2],\, 
    \Tr[\rho^2] - \Tr[\rho_B^2] \Big\} 
    \;\equiv\; 
    \mathscr{C}^{2}_{\mathrm{LB}}, \nonumber\\
    \left(\mathcal{C}(\rho)\right)^2
    &\leq 
    2\, \min\Big\{ 
    1 - \Tr[\rho_A^2],\, 
    1 - \Tr[\rho_B^2] 
    \Big\}
    \;\equiv\; 
    \mathscr{C}^{2}_{\mathrm{UB}},
\end{align}
where $\rho_A=\Tr_B(\rho)$ and $\rho_B = \Tr_A(\rho)$ are the reduced density matrices obtained by partial tracing over the degrees of freedom of the other subsystem. The positive lower bound $\mathscr{C}_{\mathrm{LB}} > 0$ guarantees the presence of entanglement, while the null upper bound $\mathscr{C}_{\mathrm{UB}} = 0$ implies separability. Throughout this work, we use $\mathscr{C}_{\mathrm{LB}}$ and $\mathscr{C}_{\mathrm{UB}}$ to assess the entanglement properties of the $WW^*$ system in Higgs decays.

\section{Quantum Tomography}
\label{sec:tomography}

The $WW$ spin system, as a bipartite state of massive spin-1 particles, is based on a nine-dimensional Hilbert space. To analyze its quantum properties, we parameterize the density matrix using a complete operator basis constructed from irreducible tensor operators. The expansion reads~\cite{Aguilar-Saavedra:2022wam,Barr:2024djo}
\begin{equation}
    \rho
    \;=\;
    \frac {1}{9}\left( \mathbb{I}_3 \otimes\mathbb{I}_3+A^{1}_{LM} T^{L}_{M}\otimes \mathbb{I}_3+ A^{2}_{LM}\mathbb{I}_3\otimes T^{L}_{M}+ C_{L_1 M_1 L_2 M_2} T^{L_1}_{M_1}\otimes T^{L_2}_{M_2}\right)\,,
\label{eq:rho}
\end{equation}
with implicit summation over the indices $L,L_{1,2}=1,2$ and $M,M_{1,2}=-L,...,L$. The operators $T^L_M$ form a complete set of normalized irreducible tensors satisfying
\begin{equation}
    (T^L_M)^\dagger = (-1)^M T^L_{-M}, \qquad \Tr\left[ T^L_M (T^L_M)^\dagger \right] = 3.
\end{equation}

We focus on the Higgs decay process $h\to WW^*\to\ell^+\nu_\ell\ell'^-\bar{\nu}_{\ell'}$. Given the mass of the Higgs boson $m_h=125.09$~GeV~\cite{ATLAS:2023oaq}, one of the $W$ bosons is necessarily off-shell. Nonetheless, it can still be treated as a spin-1 object, as the contribution from the scalar component of its propagator cancels when coupled to massless final-state fermions~\cite{Peskin:1995ev,Aguilar-Saavedra:2022wam}. For definiteness, we consider the leptonic final state $W^+ \to e^+ \nu_e$ and $W^- \to \mu^- \bar{\nu}_\mu$. The decay matrix for the $W$ boson in two leptons reads
\begin{equation}
\resizebox{\textwidth}{!}{$
    \Gamma
    \;=\;
    \frac 14\begin{pmatrix}
    1+\cos^2 \theta-2\eta_\ell \cos\theta & 
    \frac {1}{\sqrt{2}}\left( \sin 2\theta-2 \eta_\ell\sin\theta \right) e^{i \varphi} & 
    \left( 1-\cos^2 \theta \right) e^{i2 \varphi} \\
    \frac {1}{\sqrt{2}}\left( \sin 2\theta-2 \eta_\ell\sin\theta \right) e^{-i \varphi} & 
    2 \sin^2\theta & 
    -\frac {1}{\sqrt{2}}\left( \sin 2 \theta+2 \eta_\ell \sin \theta \right) e^{i \varphi} \\
    \left( 1-\cos^2 \theta \right) e^{-i2 \varphi} & 
    -\frac {1}{\sqrt{2}}\left( \sin 2 \theta+2 \eta_\ell \sin \theta \right) e^{-i \varphi}  & 
    1+\cos^2\theta + 2 \eta_\ell \cos \theta
\end{pmatrix},
$}
\label{eq:Gamma_mat}
\end{equation}
where $\theta$ and $\varphi$ are the polar and azimuthal angles of the charged lepton in its parent $W$ boson rest frame. The spin-analyzing power $\eta_\ell$ for the decay of a $W^-$ ($W^+$) boson to a negatively (positively) charged lepton, \emph{i.e.}, a fermion (antifermion), is $\eta_\ell = +1$ ($-1$), reflecting the purely left-handed structure of the $W$–lepton interaction. In this study, we adopt the helicity basis to define the lepton angles~\cite{Bernreuther:2015yna,Aguilar-Saavedra:2022wam}:
\begin{itemize}
    \item $\hat{z}$ is the direction of the $W^+$ boson in the $WW$ rest frame.
    \item $\hat{x} = \mathrm{sign}(\cos\theta_{\mathrm{CM}})(\hat{p}-\cos\theta_{\mathrm{CM}} \hat{z})/\sin\theta_{\mathrm{CM}}$, where $\hat{p}=(0,0,1)$ 
    and $\cos\theta_{\mathrm{CM}} = \hat{p}\cdot\hat{z}$.
    \item $\hat{y} = \hat{z}\times\hat{x}$.
\end{itemize}

The normalized differential decay distribution for the process $h\to WW^* \to \ell_1^+ \nu_1 \ell_2^- \bar{\nu}_2$ is then expressed as
\begin{align}
    \frac{1}{\Gamma} \frac{d\Gamma}{d\Omega_1 d\Omega_2}
    \;=\;&
    \left( \frac{3}{4\pi} \right)^2 
    \Tr\left\{ \rho \left( \Gamma_1 \otimes \Gamma_2 \right)^T \right\}
    \label{eq:dsigma_fac}\\
    \;=\; &\frac{1}{(4\pi)^2} \bigg[
    1 + A^1_{LM} B_L Y^M_L(\theta_1,\varphi_1)
      + A^2_{LM} B_L Y^M_L(\theta_2,\varphi_2)\nonumber\\
      &\hspace{1cm}+ C_{L_1 M_1 L_2 M_2} B_{L_1} B_{L_2} Y^{M_1}_{L_1}(\theta_1,\varphi_1) Y^{M_2}_{L_2}(\theta_2,\varphi_2)
    \bigg],    
\label{eq:dsigma_WW}
\end{align}
where $\Gamma_i \equiv \Gamma(\theta_i, \varphi_i)$ for $i=1,2$, and $B_1=-\sqrt{2 \pi}\eta_\ell$,  $B_2=\sqrt{{2\pi}/{5}}$. 
The angular coefficients $A_{LM}^{i}$ and $C_{L_1 M_1 L_2 M_2}$ can be determined from the orthogonality relations of the spherical harmonics:
\begin{align}
\frac {1}{\sigma} \int \frac{d \sigma}{d\Omega_1 d\Omega_2}Y_L^M(\Omega_i)^{\ast} d \Omega_1 d \Omega_2=& \frac {B_L}{4\pi}A^i_{LM},  \label{eq:Coeffs1}\\
\frac {1}{\sigma} \int \frac{d \sigma}{d\Omega_1 d\Omega_2}Y_{L_1}^{M_1}(\Omega_1)^{\ast} Y_{L_2}^{M_2}(\Omega_2)^{\ast}  d \Omega_1 d \Omega_2=&\frac {B_{L_1} B_{L_2}}{(4\pi)^2}C_{L_1 M_1 L_2 M_2}.\label{eq:Coeffs2}
\end{align}

The use of~\autoref{eq:Coeffs1} and~\autoref{eq:Coeffs2} to extract the angular coefficients guarantees the hermiticity of the spin density matrix $\rho$, as it leads to the relations $A^i_{LM} = (-1)^M (A^i_{L,-M})^*$ and $C_{L_1 M_1 L_2 M_2} = (-1)^{M_1 + M_2} (C_{L_1,-M_1, L_2,-M_2})^*$. Moreover, the expansion in~\autoref{eq:rho} enforces the proper normalization of the density matrix. However, the physical requirement that $\rho$ be positive semi-definite is not automatically ensured in this reconstruction~\cite{Goncalves:2025qem}. 
Any violation of this condition typically reflects a breakdown in the modeling assumptions, either in the two-qutrit assumption~\eqref{eq:rho}, the decay matrix~\eqref{eq:Gamma_mat}, or the factorized angular distribution~\eqref{eq:dsigma_fac}. We explore the possibility and implications of such breakdowns from NLO EW corrections for the $h\to WW^*$ decay in~\autoref{sec:analysis}.

\section{Analysis}
\label{sec:analysis}

\begin{figure}[!tb]
    \centering
    \includegraphics[width=0.8\textwidth]{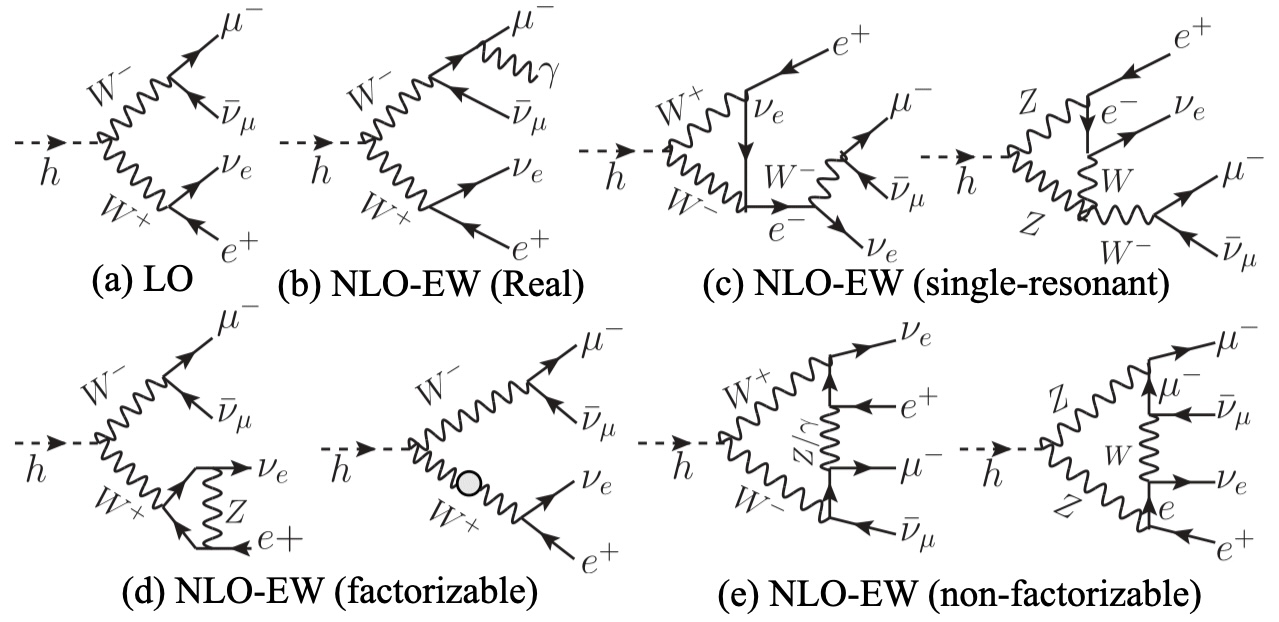}
    \caption{Representative sample of Feynman diagrams for the LO and NLO EW contributions to the Higgs boson decay into four leptons $h \rightarrow e^+ \nu_e \mu^- \bar{\nu}_{\mu}$.}
    \label{fig:Feyn}
\end{figure}

The decay channel $h \to WW^\ast \to \ell^+ \nu_\ell \ell'^- \bar{\nu}_{\ell'}$ offers a compelling framework for studying quantum entanglement in a two-qutrit system composed of massive spin-1 gauge bosons at the LHC. A representative sample of Feynman diagrams at leading and next-to-leading electroweak orders is shown in~\autoref{fig:Feyn}. Due to the scalar nature of the Higgs boson, the $WW^\ast$ system is produced in a spin-correlated state, which, at leading order, enforces strict constraints on the angular coefficients $A_{LM}^i$ and $C_{L_1 M_1 L_2 M_2}$. These constraints result in a particular texture for the spin density matrix, which implies the presence of quantum entanglement, even for inclusive final-state configurations~\cite{Barr:2021zcp,Aguilar-Saavedra:2022wam,Fabbrichesi:2023cev,Fabbrichesi:2023jep,Fabbri:2023ncz,Bernal:2023ruk,Aguilar-Saavedra:2024whi,Subba:2024mnl,Bernal:2024xhm,Sullivan:2024wzl,Aguilar-Saavedra:2024jkj}. In what follows, we first review the leading-order angular coefficients $A_{LM}^i$ and $C_{L_1 M_1 L_2 M_2}$, and quantify the degree of entanglement for the $WW^\ast$ system. We then derive how higher-order electroweak corrections modify these observables, and assess their effects on the density matrix and two-qutrit interpretation.

\subsection{Leading Order Decay}

In the decay $h \to WW^\ast$, angular momentum and parity conservation impose strong constraints on the spin correlations, resulting in the following relations among the angular coefficients at LO~\cite{Aguilar-Saavedra:2022wam,DelGratta:2025qyp}:
\begin{align}
    A^{1}_{20} &= A^{2}_{20} \neq 0,\nonumber\\
    C_{1,1,1,-1} &= C_{1,-1,1,1} = -C_{2,1,2,-1} = -C_{2,-1,2,1} \neq 0,\label{eq:LOConds}\\
    C_{2,2,2,-2} &= C_{2,-2,2,2} = -C_{1,0,1,0} = 2 - C_{2,0,2,0} \neq 0,\nonumber\\
    \frac{A^{1,2}_{20}}{\sqrt{2}} + 1 &= C_{2,2,2,-2}.\nonumber
\end{align}
These conditions constrain the LO density matrix to the following form:
\begin{align}
    \label{eq:rhoLOForm}
    \rho_{\mathrm{LO}} = \begin{pmatrix}
        0 & 0 & 0 & 0 & 0 & 0 & 0 & 0 & 0 \\
        0 & 0 & 0 & 0 & 0 & 0 & 0 & 0 & 0 \\
        0 & 0 & \frac{1}{3}C_{2,2,2,-2} & 0 & \frac{1}{3}C_{2,1,2,-1} & 0 & \frac{1}{3}C_{2,2,2,-2} & 0 & 0 \\
        0 & 0 & 0 & 0 & 0 & 0 & 0 & 0 & 0 \\
        0 & 0 & \frac{1}{3}C_{2,1,2,-1} & 0 & \frac{1}{3}(3 - 2C_{2,2,2,-2}) & 0 & \frac{1}{3}C_{2,1,2,-1} & 0 & 0 \\
        0 & 0 & 0 & 0 & 0 & 0 & 0 & 0 & 0 \\
        0 & 0 & \frac{1}{3}C_{2,2,2,-2} & 0 & \frac{1}{3}C_{2,1,2,-1} & 0 & \frac{1}{3}C_{2,2,2,-2} & 0 & 0 \\
        0 & 0 & 0 & 0 & 0 & 0 & 0 & 0 & 0 \\
        0 & 0 & 0 & 0 & 0 & 0 & 0 & 0 & 0 \\
    \end{pmatrix},
\end{align}
with only nine nonzero entries, all fully determined by two independent parameters, denoted here as $C_{2,2,2,-2}$ and $C_{2,1,2,-1}$.

To numerically estimate these coefficients and density matrix, we generate events for $h\to WW^\ast\to e^+ \nu_e \mu^- \bar{\nu}_{\mu}$ at LO with the \texttt{Prophecy4F} package~\cite{Denner:2019fcr}. We adopt the following on-shell input parameters:
\begin{align}
    m_h &= 125~\mathrm{GeV}, && \Gamma_h = 4.097~\mathrm{MeV}, \label{eq:mh}\\
    m_t &= 173.2~\mathrm{GeV}, && \Gamma_t =  1.369~\mathrm{GeV}, \label{eq:mt} \\
    m_W &=  80.385~\mathrm{GeV}, && \Gamma_W = 2.085~\mathrm{GeV}, \label{eq:CMparams1}\\
    m_Z &=  91.1876~\mathrm{GeV}, && \Gamma_Z =  2.4952~\mathrm{GeV}, \label{eq:CMparams2}\\
    G_\mu &= 1.1663787\times10^{-5}~\mathrm{GeV}^{-2}. \label{eq:GFparam}&&   
\end{align}
The EW coupling is fixed with the $G_\mu$ scheme and the complex mass scheme is used for the treatment of unstable particles~\cite{Denner:1999gp,Denner:2005fg}. We then follow the quantum tomography formalism presented in~\autoref{sec:tomography}, extracting the angular coefficients using~\autoref{eq:Coeffs1} and~\autoref{eq:Coeffs2}.\footnote{Rather than focusing on the neutrino reconstruction, which can be performed using established methods~\cite{Cho:2008tj,Barr:2010zj,CMS:2017znf,Goncalves:2018agy,ATLAS:2022ooq,CMS:2022uhn,Ackerschott:2023nax,CMS:2024bua,Dong:2024fzt,ATLAS:2025abg}, we investigate how the underlying angular coefficients are affected by higher-order corrections, assuming perfect knowledge of the neutrino momenta. These coefficients form the theoretical benchmark that reconstruction techniques would ultimately aim to approximate.} 
The resulting density matrix reads
\begin{align}
\rho_{\mathrm{LO}} = \begin{pmatrix}
            0 & 0 & 0 & 0 & 0 & 0 & 0 & 0 & 0 \\
            0 & 0 & 0 & 0 & 0 & 0 & 0 & 0 & 0 \\
            0 & 0 & 0.1972(2) & 0 & -0.3139(3) & 0 & 0.1962(5) & 0 & 0 \\ 
            0 & 0 & 0 & 0 & 0 & 0 & 0 & 0 & 0 \\
            0 & 0 & -0.3139(3) & 0 & 0.6037(4) & 0 & -0.3137(3) & 0 & 0 \\
            0 & 0 & 0 & 0 & 0 & 0 & 0 & 0 & 0 \\
            0 & 0 & 0.1962(5) & 0 & -0.3137(3) & 0 & 0.1968(2) & 0 & 0 \\
            0 & 0 & 0 & 0 & 0 & 0 & 0 & 0 & 0 \\
            0 & 0 & 0 & 0 & 0 & 0 & 0 & 0 & 0 \\
    \end{pmatrix}\,,
        \label{eq:hWWdensLO}
\end{align}
where the zero entries indicate compatibility with zero within numerical uncertainties. A comparison reveals that the structure of the density matrix obtained numerically in~\autoref{eq:hWWdensLO} is consistent with the analytical form of~\autoref{eq:rhoLOForm}.

\begin{figure}[!tb]
    \centering
    \includegraphics[width=0.5\textwidth]{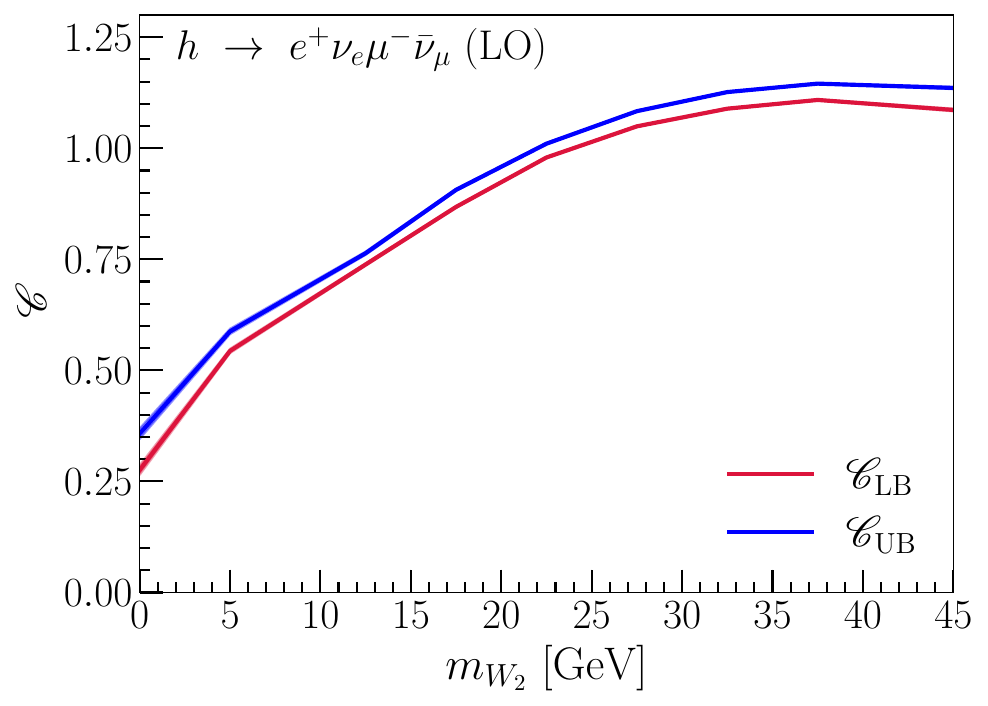}
    \caption{Lower $\mathscr{C}_{\mathrm{LB}}$ (red line) and upper $\mathscr{C}_{\mathrm{UB}}$ (blue line) bounds of the concurrence for $h\to WW^\ast\to e^+ \nu_e \mu^- \bar{\nu}_{\mu}$ at LO as a function of the lowest reconstructed invariant mass, $m_{W_2}<m_{W_1}$.}
    \label{fig:CboundsLO}
\end{figure}

In \autoref{fig:CboundsLO}, we present the leading order bounds on the concurrence, $\mathscr{C}_{\mathrm{LB}}$ and $\mathscr{C}_{\mathrm{UB}}$, as a function of the reconstructed mass of the subleading $W$ boson, $m_{W_2}$\footnote{We used the publicly available Python package \texttt{pyerrors}~\cite{Joswig:2022qfe}, which is based on the $\Gamma$-method~\cite{Wolff:2003sm, Ramos:2018vgu, Ramos:2020scv} to propagate the uncertainties from the density matrix to its eigenvalues and concurrence bounds.}. The fact that both bounds remain strictly positive across the entire mass range signals the presence of entanglement. Furthermore, the concurrence increases with $m_{W_2}$, approaching its maximal value as the $WW^\ast$ system tends toward a pure state~\cite{Aguilar-Saavedra:2022wam}.

\begin{figure}[!b]
    \centering
    \includegraphics[width=0.3\textwidth]{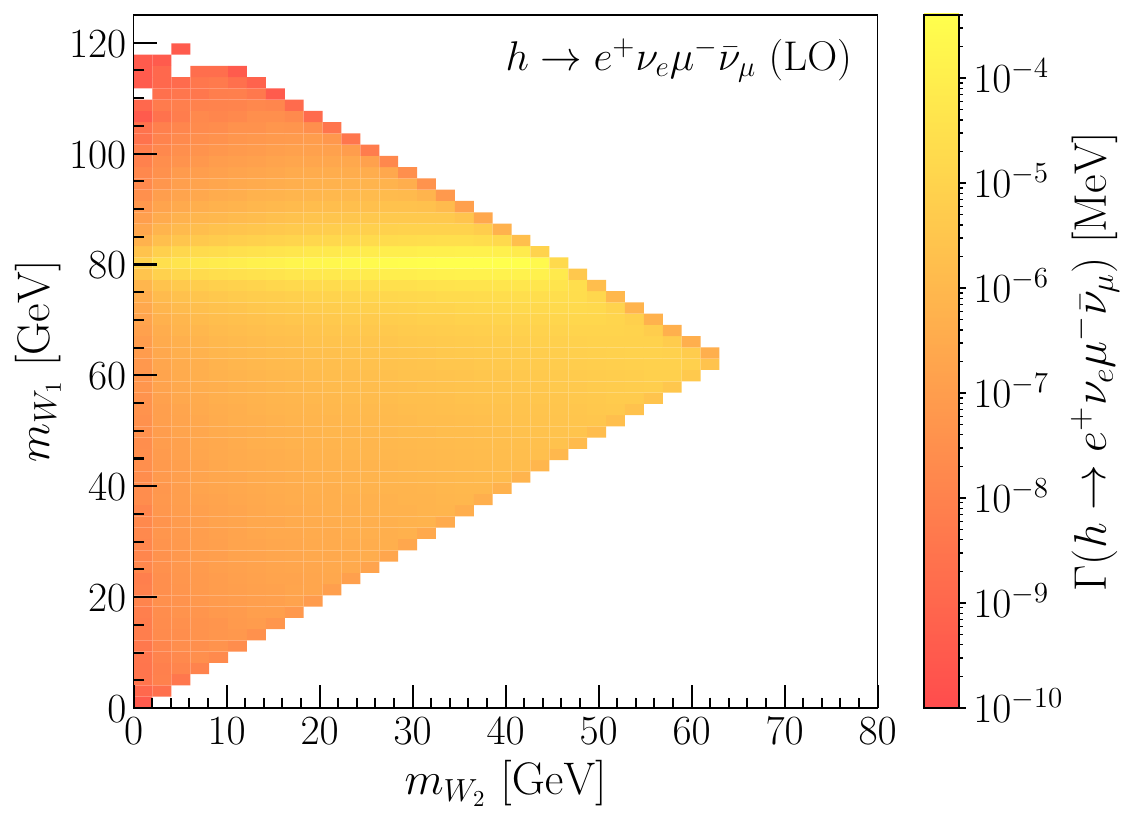}
    \includegraphics[width=0.3\textwidth]{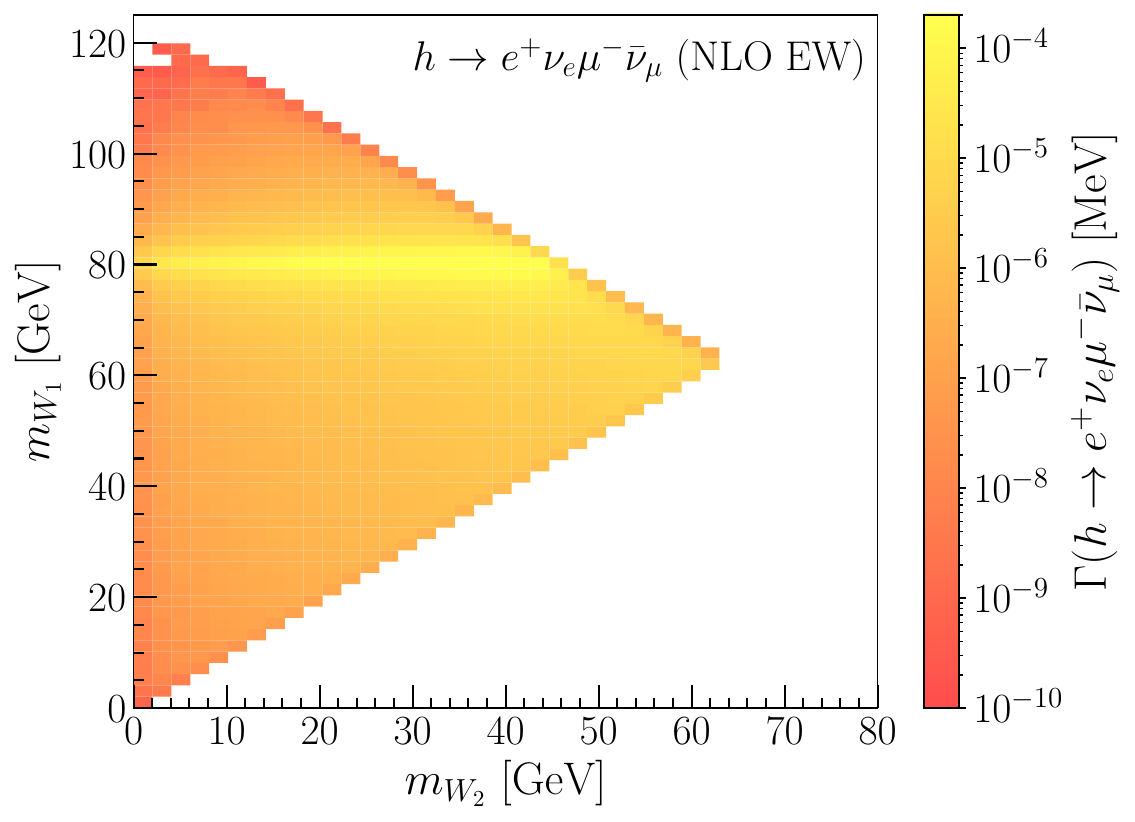}
    \includegraphics[width=0.3\textwidth]{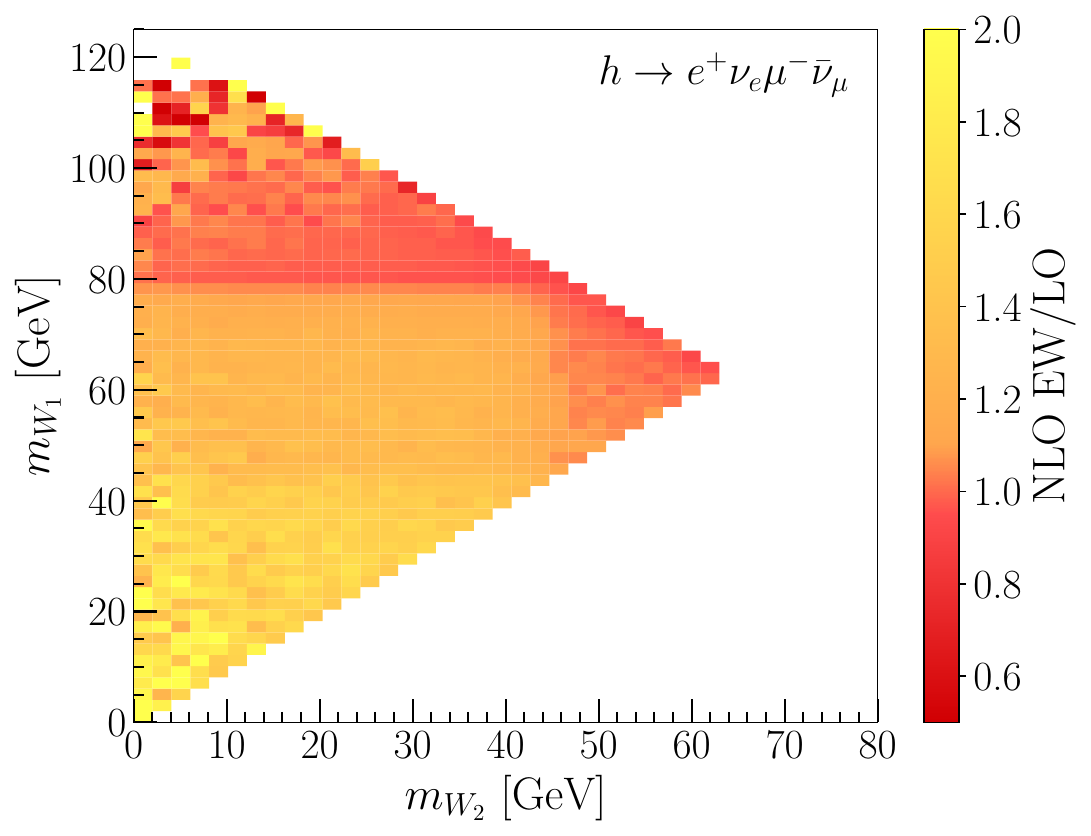}
    \caption{Decay distribution for $h \rightarrow e^+ \nu_e \mu^- \bar{\nu}_{\mu}$ at LO (left panel), NLO EW (central panel), and their corresponding NLO/LO ratios (right panel) in the $m_{W_{1}}$-$m_{W_{2}}$ plane.}
    \label{fig:Width}
\end{figure}

\subsection{NLO EW Effects}

To evaluate the NLO electroweak (EW) effects, we use \texttt{Prophecy4F} with the same input parameters as in the LO analysis, see \cref{eq:mh,eq:mt,eq:CMparams1,eq:CMparams2,eq:GFparam}, within the complex mass scheme~\cite{Denner:1999gp,Denner:2005fg,Bredenstein:2006rh}. Photon radiation is included by dressing charged leptons within a cone of $\Delta R(\ell, \gamma) < 0.1$.
In~\autoref{fig:Width}, we present the partial decay width for the process $h \rightarrow e^+ \nu_e \mu^- \bar{\nu}_\mu$ at LO and NLO EW, along with the corresponding NLO/LO ratio, shown as a function of the highest and lowest reconstructed invariant masses of the $\ell \nu_\ell$ pairs, denoted as $m_{W_1}$ and $m_{W_2}$, respectively. We assume perfect reconstruction of the neutrino momenta, focusing on the higher-order effects. The distributions reveal that the dominant contribution to the partial Higgs decay width arises when the leading mass is near the $W$ pole, $|m_{W_1} - m_W| \lesssim 10~\text{GeV}$, while the subleading mass lies in the range $10~\text{GeV} \lesssim m_{W_2} \lesssim 40~\text{GeV}$. In this region, the NLO/LO ratio remains close to unity, indicating mild higher-order corrections. This moderate behavior is also reflected in the integrated NLO EW correction to the partial width, with $\delta\Gamma_\text{NLO}/\Gamma_\text{LO} \simeq 3\%$.

\begin{table}[!t]
    \centering
    \begin{tabular}{|c|c|c|c|}
       \hline Coefficient & LO & NLO EW & NLO/LO \\ \hline
        $A_{2,0}^{1}$ & $-0.5769(6)$ & $-0.550(1)$ & $0.953$ \\
        $A_{2,0}^{2}$ & $-0.5763(6)$ & $-0.550(1)$ & $0.954$ \\
        $C_{1,0,1,0}$ & $-0.5917(3)$ & $-0.6034(4)$ & $1.0198$ \\
        $C_{1,1,1,-1}$ & $0.9439(3)$ & $0.9429(3)$ & $0.9989$ \\
        $C_{2,1,2,-1}$ &  $-0.939(1)$ & $-0.949(2)$ & $0.989$ \\
        $C_{2,0,2,0}$ & $1.401(2)$ & $1.383(2)$ & $0.987$ \\
        $C_{2,2,2,-2}$ & $0.5885(5)$ & $0.590(1)$ & $1.002$ \\
        $A_{1,0}^{1}$ & $0$ & $-0.0122(3)$ & $-$ \\
        $A_{1,0}^{2}$ & $0$ & $-0.0115(3)$ & $-$ \\
        $C_{1,-1,2,1}$ & $0$ & $-0.0153(6)$ & $-$ \\
        $C_{1,0,2,0}$ & $0$ & $0.0244(7)$ & $-$ \\
        $C_{2,-1,1,1}$ & $0$ & $-0.0169(6)$ & $-$ \\
        $C_{2,0,1,0}$ & $0$ & $0.0232(7)$ & $-$ \\
        \hline 
    \end{tabular}
    \caption{The non-vanishing coefficients for the $h\to  e^+ \nu_e \mu^- \bar{\nu}_{\mu}$ decay at LO, NLO EW, and their NLO/LO ratio. The index $i=1,2$, as used for example in $A^i_{LM}$, denotes the polarization of $W^+$ and $W^-$ bosons, respectively.    
    Additional non-vanishing $C_{L_1,M_1,L_2,M_2}$ coefficients can be obtained from those shown using the relation $C_{L_1,M_1,L_2,M_2} = (-1)^{M_1 + M_2} (C_{L_1,-M_1, L_2,-M_2})^*$, which follows from the symmetry properties of spherical harmonics. Charged leptons are dressed with photons using a recombination radius of $\Delta R(\ell,\gamma)<0.1$. All coefficients have vanishing imaginary part within uncertainties.}
    \label{tab:indcoeffsLONLO}
\end{table}

We now investigate the impact of higher-order corrections on the angular coefficients. In~\autoref{tab:indcoeffsLONLO}, we present the non-vanishing coefficients, within uncertainties, for the decay \( h \to e^+ \nu_e \mu^- \bar{\nu}_{\mu} \) at LO, NLO EW, and their corresponding NLO/LO ratio. Systematic uncertainties due to the missing two-loop EW contribution are estimated by taking the square of the known one-loop corrections~\cite{Freitas:2019bre}. The uncertainties shown in parentheses account for both statistical and systematic errors propagated from the angular distributions.  We observe that the LO relations given by~\autoref{eq:LOConds} are visibly violated at NLO. This is reflected in corrections of up to \(\sim 5\%\) to previously dominant coefficients, as well as the emergence of novel angular coefficients that vanish at LO, enriching the angular dynamics.

Angular observables provide sensitive probes of physics beyond the Standard Model. To draw robust conclusions, it is crucial to evaluate these observables with high precision, including higher-order corrections. However, the interpretation of the process as a two-qutrit system, and consequently the applicability of the quantum tomography framework described in~\autoref{sec:tomography}, becomes questionable once NLO electroweak corrections are considered. Several types of radiative corrections challenge this interpretation. For instance, virtual contributions, such as the single-resonant topology illustrated in~\autoref{fig:Feyn}~(c) and the non-factorizable diagrams in~\autoref{fig:Feyn}~(e), indicate that the two-qutrit system defined in~\autoref{eq:rho} and the factorized form of the cross section in~\autoref{eq:dsigma_fac} may no longer hold. Moreover, the presence of real photon emission, depicted in~\autoref{fig:Feyn}~(b), when the photon is not recombined, alters the form of the decay matrix from~\autoref{eq:Gamma_mat}. These radiative effects call into question the applicability of the idealized tomography formalism in~\autoref{sec:tomography}, suggesting that the interpretation of the $WW^*$ system as a two-qutrit quantum state may break down once full NLO corrections are taken into account.

To assess the numerical impact of higher-order effects on the validity of the two-qutrit interpretation, we apply the quantum tomography procedure described in~\autoref{sec:tomography} and examine whether the fundamental properties of the reconstructed density matrix are preserved. The resulting matrix, including full NLO electroweak corrections, takes the form
\begin{align}
\resizebox{0.95\textwidth}{!}{$
\rho_{\mathrm{NLO}} = \begin{pmatrix}
            0.0023(2) & 0 & 0 & 0 & 0 & 0 & 0 & 0 & 0 \\
            0 & -0.0056(3) & 0 & 0.0065(3) & 0 & 0 & 0 & 0 & 0 \\
            0 & 0 & 0.2021(2) & 0 & -0.3152(3) & 0 & 0.1966(5) & 0.0011(3) & 0 \\ 
            0 & 0.0065(3) & 0 & -0.0054(3) & 0 & 0 & 0 & 0 & 0 \\
            0 & 0 & -0.3152(3) & 0 & 0.5913(5) & 0 & -0.3157(3) & 0 & 0 \\
            0 & 0 & 0 & 0 & 0 & 0.0066(3) & 0 & -0.0042(3) & 0 \\
            0 & 0 & 0.1966(5) & 0 & -0.3157(3) & 0 & 0.2021(2) & 0 & 0 \\
            0 & 0 & 0.0011(3) & 0 & 0 & -0.0042(3) & 0 & 0.0070(3) & 0 \\
            0 & 0 & 0 & 0 & 0 & 0 & 0 & 0 & 0 \\
    \end{pmatrix},
    $}
    \label{eq:hWWdensNLO}
\end{align}
where the vanishing entries are consistent with zero within $1\sigma$ uncertainties. Compared to the LO case shown in~\cref{eq:rhoLOForm,eq:hWWdensLO}, the NLO corrections introduce both modifications to existing angular coefficients and entirely new nonzero components, resulting in a new texture for the density matrix at NLO. As remarked in~\autoref{sec:tomography}, the presented quantum tomography enforces hermiticity and normalization of the density matrix. However, it does not automatically guarantee that the matrix is positive semi-definite, a critical physical requirement. To test this condition, we evaluate the eigenvalues of $\rho_\text{NLO}$ and find that the smallest eigenvalue is negative, $-0.0122(5)$, with another negative eigenvalue of significantly smaller magnitude, $-0.0007(3)$. These results formally signal the breakdown of the two-qutrit interpretation due to higher-order effects.  

\begin{figure}[!tb]
    \centering
    \includegraphics[width=0.3\textwidth]{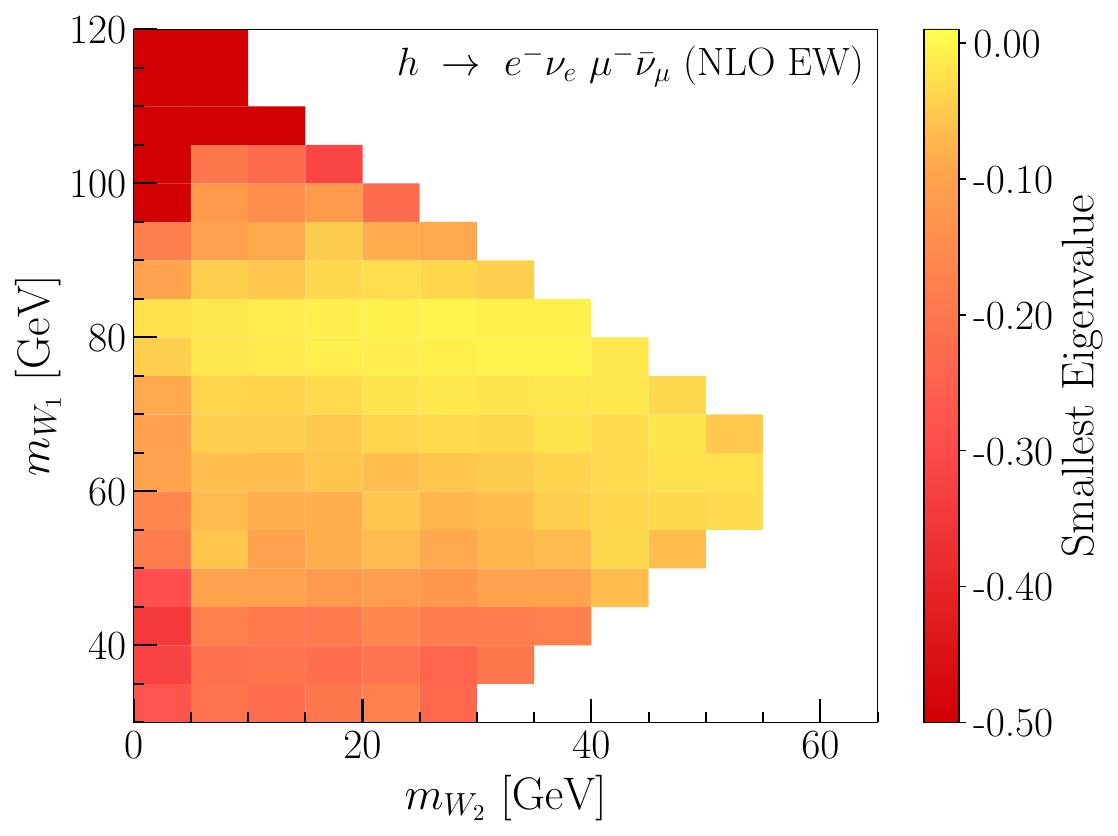}
     \includegraphics[width=0.3\textwidth]{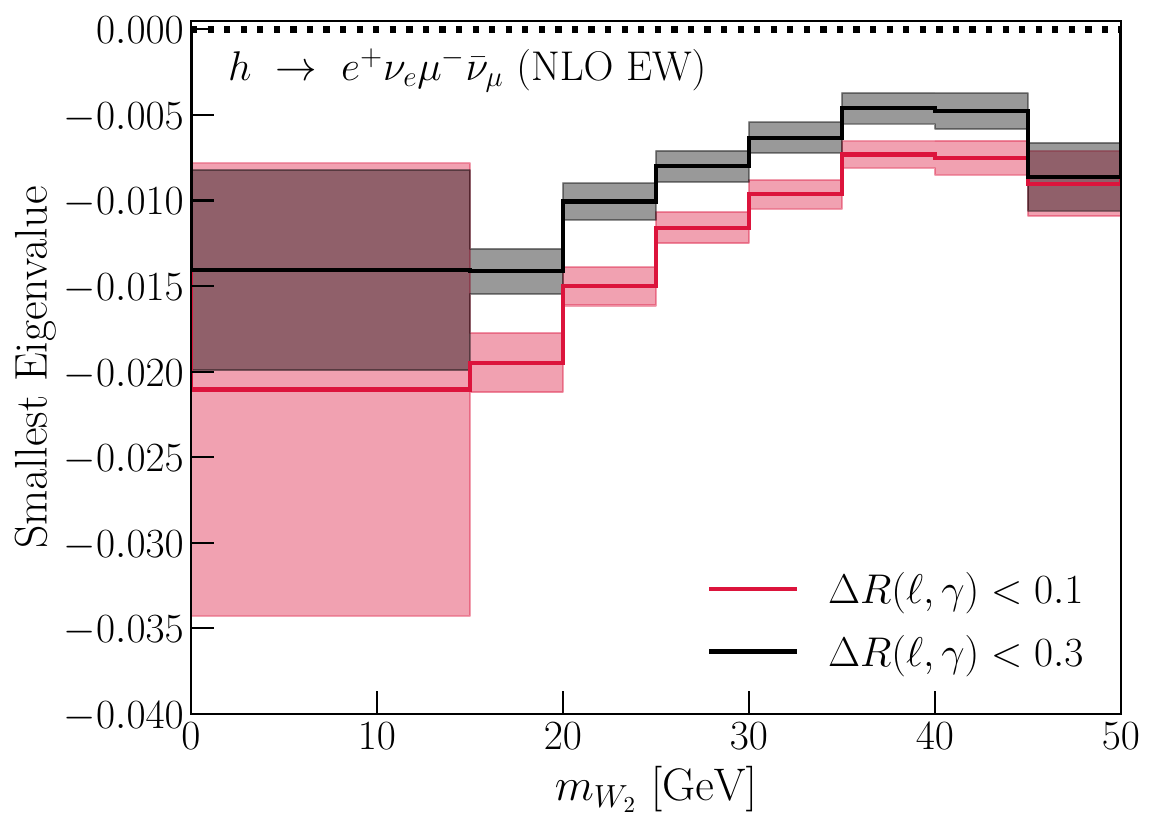}
     \includegraphics[width=0.3\textwidth]{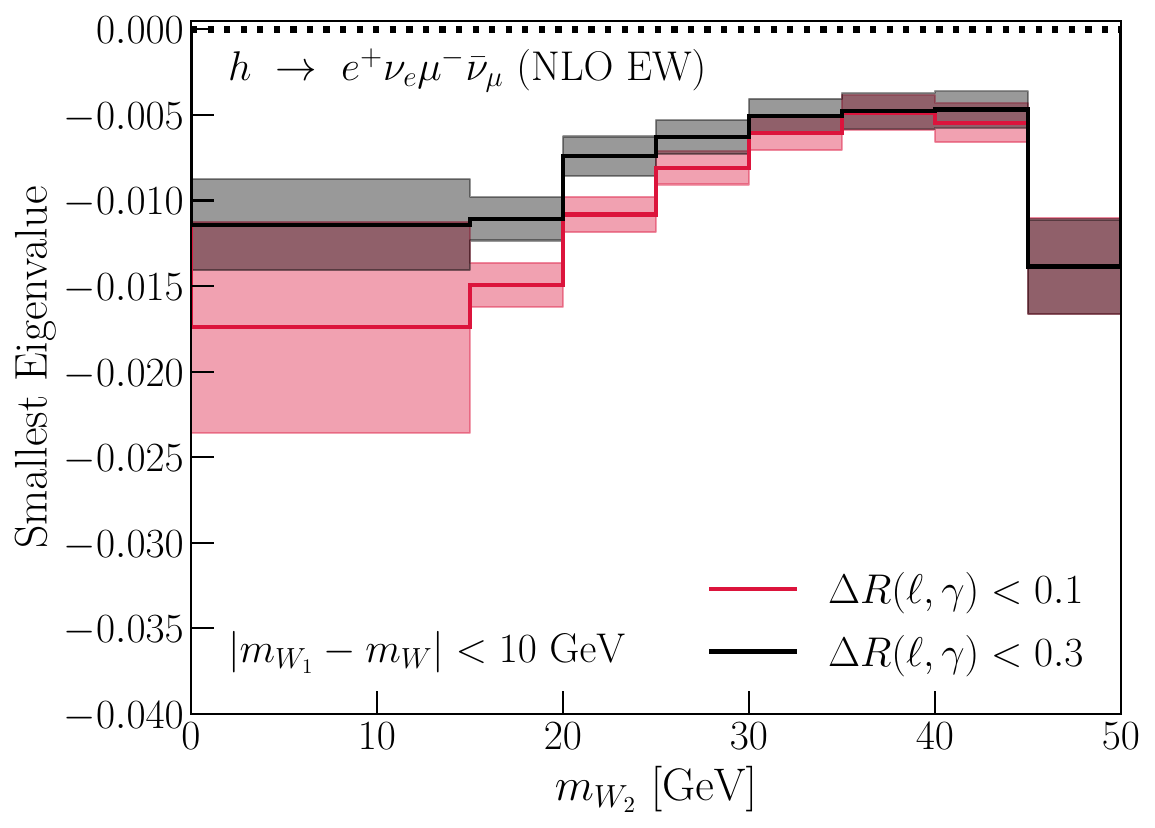}
    \caption{Smallest eigenvalue of the density matrix $\rho_{\mathrm{NLO}}$ as a function of the two-dimensional distribution $(m_{W_2}, m_{W_1})$. The central panel shows the same quantity as a function of $m_{W_2}$ and the right panel displays it after imposing a cut on the reconstructed mass closest to the $W$ boson pole, $|m_{W_1} - m_W| < 10$~GeV.  The left panel assumes the photon-lepton recombination criterion $\Delta R(\ell,\gamma)<0.1$, whereas the central and right panels display results for two possible criteria: $\Delta R(\ell,\gamma)<0.1$ (red) and $\Delta R(\ell,\gamma)<0.3$ (black).}
    \label{fig:SmallEig}
\end{figure}

While the negative eigenvalues are already very small, we investigate strategies to further suppress the undesired contributions and potentially restore a more robust two-qutrit description of the $h\to e^+ \nu_e \mu^- \bar{\nu}_{\mu}$ decay at NLO. In~\autoref{fig:SmallEig} (left panel), we present the lowest eigenvalue of the NLO density matrix as a function of the two-dimensional distribution $(m_{W_2}, m_{W_1})$. Whereas the smallest eigenvalue takes on negative values across the phase space, it approaches zero when the leading reconstructed mass, $m_{W_1}$, lies close to the $W$ boson pole. This kinematic configuration suppresses the non-factorizable contributions, such as those illustrated in~\autoref{fig:Feyn}~(e), thus improving the validity of the two-qutrit description. Fortunately, this can be achieved with minimal impact on the number of events, as this kinematic regime displays the bulk of the $h\to  e^+ \nu_e \mu^- \bar{\nu}_{\mu}$ events, as shown in~\autoref{fig:Width}. For more details on the angular coefficients in the scenario where $m_{W_1}$ is selected close to the $W$ boson pole, $|m_{W_1} - m_W| < 10$~GeV, see~\autoref{tab:indcoeffsLONLO2}. 
Another relevant kinematic handle is the dependence on the invariant mass $m_{W_2}$. The central and right panels of~\autoref{fig:SmallEig} show a pronounced dependence of the lowest eigenvalue with $m_{W_2}$, reaching values of order $\mathcal{O}(10^{-3})$ in the range $30~\text{GeV}\lesssim m_{W_2}\lesssim45$~GeV. 
Further improvements can be achieved by increasing the radius in the photon-lepton recombination criterion, as illustrated in~\autoref{fig:SmallEig} (central and right panels), where the recombination radius is varied from $\Delta R(\ell,\gamma)<0.1$ to 0.3. A larger recombination radius mitigates the impact of events in which the photon does not cluster with its parent lepton, events that are not accurately described by the decay matrix in~\autoref{eq:Gamma_mat}. 

\begin{table}[!t]
    \centering
    \begin{tabular}{|c|c|c|c|}
       \hline Coefficient & LO & NLO EW  & NLO/LO \\ \hline
        $A_{2,0}^{1}$ & $-0.5284(6)$ & $-0.5120(6)$ & $0.9689$ \\
        $A_{2,0}^{2}$ & $-0.5276(6)$ & $-0.5122(6)$ & $0.9708$ \\
        $C_{1,0,1,0}$ & $-0.6261(4)$ & $-0.6350(5)$ & $1.0142$ \\
        $C_{1,1,1,-1}$ & $0.9777(4)$ & $0.9790(2)$ & $1.0013$ \\
        $C_{2,1,2,-1}$ &  $-0.973(1)$ & $-0.983(2)$ & $1.010$ \\
        $C_{2,0,2,0}$ & $1.367(2)$ & $1.359(2)$ & $0.994$ \\
        $C_{2,2,2,-2}$ & $0.623(2)$ & $0.629(2)$ & $1.009$ \\
        $A_{1,0}^{1}$ & $0$ & $-0.0074(3)$ & $-$ \\
        $A_{1,0}^{2}$ & $0$ & $-0.0067(3)$ & $-$ \\
        $C_{1,-1,2,1}$ & $0$ & $-0.0105(7)$ & $-$ \\
        $C_{1,0,2,0}$ & $0$ & $0.0138(9)$ & $-$ \\
        $C_{2,-1,1,1}$ & $0$ & $-0.0117(8)$ & $-$ \\
        $C_{2,0,1,0}$ & $0$ & $0.0129(9)$ & $-$ \\
        $C_{1,-1,1,0}$ & $0$ & $0.0010(3)$ & $-$ \\
        $C_{2,-2,2,1}$ & $0$ & $0.005(1)$ & $-$ \\
        \hline 
    \end{tabular}
    \caption{The non-vanishing coefficients for the $h\to  e^+ \nu_e \mu^- \bar{\nu}_{\mu}$ decay at LO, NLO EW, and their NLO/LO ratio. The invariant mass conditions $|m_{W_1}-m_W|<10$~GeV and $m_{W_2}>10$~GeV are imposed. Charged leptons are dressed with photons using a recombination radius of $\Delta R(\ell,\gamma)<0.1$. All coefficients have vanishing imaginary part within the uncertainties.}
    \label{tab:indcoeffsLONLO2}
\end{table}

\begin{figure}[!tb]
    \centering
    \includegraphics[width=0.45\textwidth]{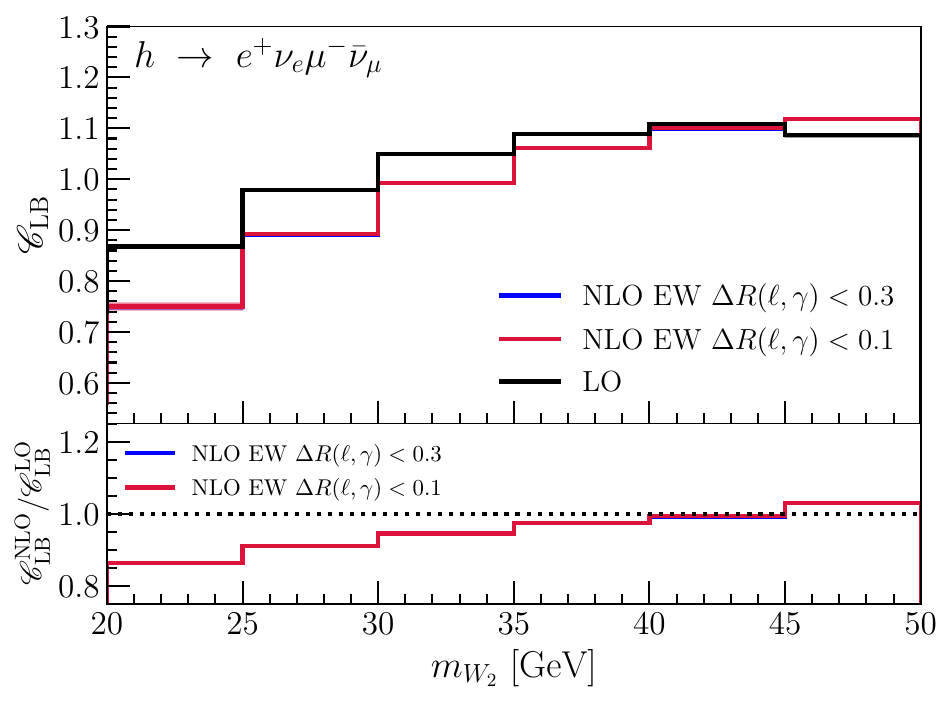}
     \includegraphics[width=0.45\textwidth]{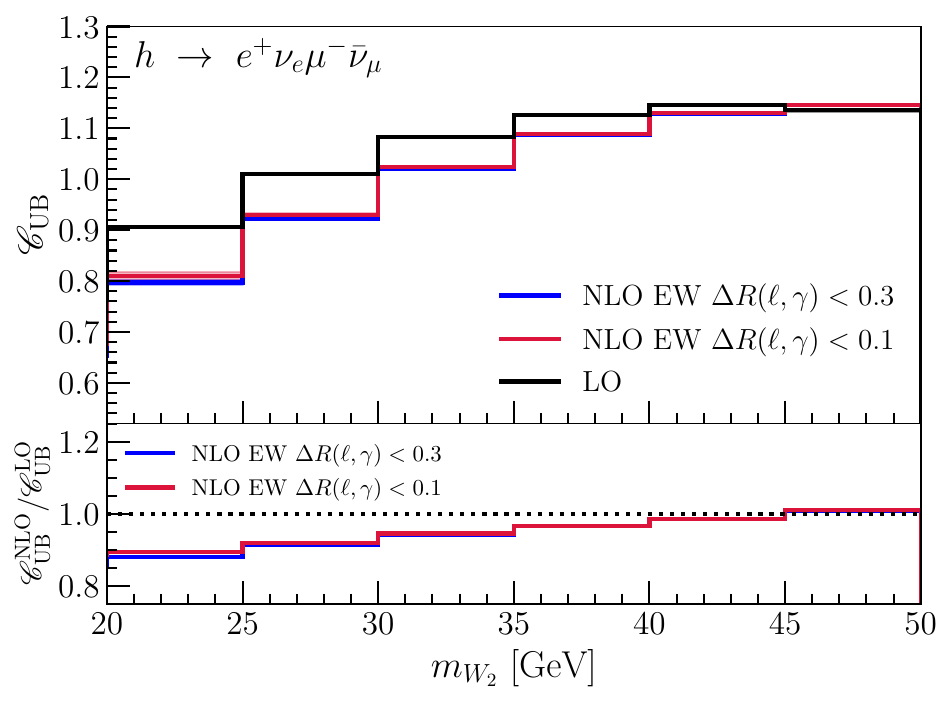}
    \caption{Lower $\mathscr{C}_{\mathrm{LB}}$ (left panel) and upper $\mathscr{C}_{\mathrm{UB}}$ (right panel) bounds of the concurrence for $h\to WW^\ast\to e^+ \nu_e \mu^- \bar{\nu}_{\mu}$ as a function of the off-shell $W$ mass, $m_{W_2}$. The results are presented at LO (black) and NLO EW with recombination radii $\Delta R(\ell,\gamma)<0.1$ (red) and $\Delta R(\ell,\gamma)<0.3$ (blue).}
    \label{fig:CboundsNLO}
\end{figure}

While negative eigenvalues are formally present, their numerical impact is subleading. Hence, the fundamental properties of the two-qutrit density matrix remain satisfied to a very good approximation at NLO EW, justifying the evaluation of the entanglement measure of the two-qutrit system at this level of precision. In~\autoref{fig:CboundsNLO}, we present the lower and upper bounds of the concurrence at LO and NLO EW.\footnote{Since the negative eigenvalues are comparatively small in magnitude, we neglect their existence when evaluating the concurrence bounds. The validity of this approximation is examined in~\autoref{app:proj}, where we compare these results with those obtained from the projected density matrix, defined as the closest positive semi-definite Hermitian matrix with unit trace. The distance between the matrices is defined by the Frobenius norm~\cite{10.5555/248979}.} For consistency, we restrict the analysis to the region where negative eigenvalues remain below $\mathcal{O}(10^{-2})$, which corresponds to $m_{W_2}\gtrsim 20~\text{GeV}$. We find that higher-order corrections yield a qualitatively similar entanglement pattern to the LO prediction, with a moderate degradation of at most 10\% at low $m_{W_2}$, and a closer agreement with the LO behavior at higher values of $m_{W_2}$. We perform the NLO EW analysis for two photon–lepton recombination scenarios: $\Delta R(\ell,\gamma) < 0.1$ (red line) and $\Delta R(\ell,\gamma) < 0.3$ (blue line). The differences between these choices in the lower and upper bounds of the entanglement measure are found to be subleading.

\subsection{Comparison with the $h \to  e^+e^-\mu^+\mu^-$ Channel}

It is instructive to compare the present analysis of NLO electroweak effects to angular coefficients in the $h \to WW^*\to e^+ \nu_e \mu^- \bar{\nu}_\mu$ channel with the corresponding study of the $h \to ZZ^*\to e^+e^-\mu^+\mu^-$ decay~\cite{Grossi:2024jae,DelGratta:2025qyp,Goncalves:2025qem}. In the $h\to e^+ \nu_e \mu^- \bar{\nu}_\mu$ channel, the absence of electric charge on neutrinos eliminates entire classes of photon-exchange loop topologies, such as those connecting neutrino lines or neutrinos to charged leptons, and also suppresses final-state photon radiation. Furthermore, the fixed helicity correlation between charged leptons and neutrinos from $W$ decay constrains interference with loop diagrams, since helicity flips require mass-suppressed terms. These factors strongly reduce loop-induced angular structures. Conversely, the $h\to e^+e^-\mu^+\mu^-$ channel involves only charged leptons, allowing photons to connect any pair of external legs. Moreover, the presence of both left- and right-handed couplings at LO leads to a rich interference pattern at NLO, generating nontrivial loop-induced structures that significantly modify angular coefficients. 

These general expectations are confirmed numerically. In the four charged lepton final state, the same angular coefficients that are nonzero at LO receive sizeable NLO EW corrections. Additionally, the relations among the coefficients that define the LO density matrix structure are significantly broken. This leads to new nonzero density matrix elements (that vanish at LO) with magnitudes comparable to the leading ones. These effects substantially impact the texture of the density matrix, as evidenced by the comparison between the LO and NLO results presented in Eqs.~(4.7) and~(4.8) of Ref.~\cite{Goncalves:2025qem} (see also Table~VI therein). As a result, the smallest eigenvalue of the density matrix in that case remained negative, reaching values as low as $\sim -0.2$ even under optimal kinematic selections, challenging the two-qutrit interpretation of the system~\cite{Goncalves:2025qem}.

In contrast, for the $h \to e^+ \nu_e \mu^- \bar{\nu}_\mu$ channel, NLO effects on the angular coefficients that are nonzero at LO are moderate. New coefficients that vanish at LO appear with small values, as shown in~\autoref{tab:indcoeffsLONLO} and~\autoref{tab:indcoeffsLONLO2}.\footnote{We have further investigated the impact of real photon emission by computing the angular coefficients for different recombination radii, $\Delta R(\ell,\gamma)$, as detailed in~\autoref{app:deltaR}.} The overall distortions in the density matrix are milder, with negative eigenvalues suppressed by roughly two orders of magnitude compared to the four charged lepton channel, see~\autoref{fig:SmallEig}. This indicates that the two-qutrit structure is much better preserved in the $h \to e^+ \nu_e \mu^- \bar{\nu}_\mu$ channel, making it more robust against non-factorizable contributions from higher-order effects.
For more details on the consistency of the density matrix in both channels, see~\autoref{app:proj}.

\section{Summary}
\label{sec:summary}

The Higgs boson decay $h \to WW^* \to \ell^+ \nu_\ell \ell'^- \bar{\nu}_{\ell'}$ provides a unique window into the structure of Higgs boson couplings to electroweak gauge bosons. These interactions manifest in the angular distributions of the final-state leptons, making the associated angular coefficients from gauge boson decays powerful observables for testing SM predictions and uncovering potential signs of new physics. More recently, a novel and complementary perspective has emerged by interpreting these coefficients through the framework of quantum information theory, such as by assessing whether the gauge bosons exhibit quantum entanglement, enriching the phenomenological relevance of such observables.

In this work, we have presented a systematic study of higher-order electroweak corrections to the angular coefficients in the $h \to WW^* \to \ell^+ \nu_\ell \ell'^- \bar{\nu}_{\ell'}$ channel. At leading order, the angular coefficients for the $h\to \ell^+ \nu_\ell \ell'^- \bar{\nu}_{\ell'}$ system are highly constrained by the spin and parity of the Higgs boson, yielding a density matrix with only nine non-vanishing entries, governed by just two independent parameters. This compact structure leads to a physically consistent, entangled two-qutrit quantum system, even when no phase-space cuts are applied. When NLO electroweak corrections are included, this structure is deformed. 
New angular coefficients acquire nonzero values and the previous relations among angular coefficients are broken. These deviations arise from various radiative effects, including singly-resonant topologies, non-factorizable contributions, and real photon emissions in the final state. 

A particularly illuminating comparison can be drawn between the present analysis of NLO electroweak corrections to angular coefficients in $h \to WW^*\to e^+ \nu_e \mu^- \bar{\nu}_\mu$ and the corresponding study of $h \to ZZ^*\to e^+e^- \mu^+ \mu^-  $ presented in Ref.~\cite{Goncalves:2025qem}. We find that the angular coefficients in the $h \to e^+ \nu_e \mu^- \bar{\nu}_\mu$ decay exhibit greater resilience to radiative effects than the $h  \to e^+e^- \mu^+ \mu^-$ counterpart. In the latter, NLO EW corrections, especially those arising from singly-resonant topologies, significantly distort the angular coefficients and undermine the validity of the two-qutrit interpretation. In contrast, the $h\to WW^*$ channel displays milder corrections, better preserving the two-qutrit system. 

In conclusion, our results highlight the importance of accounting for higher-order electroweak effects when studying quantum observables at colliders. While leading-order analyses offer valuable theoretical guidance, they may fail to capture the full quantum structure of realistic events. Incorporating radiative corrections is essential not only for achieving precision in Higgs coupling measurements, but also for enabling a robust interpretation of emergent quantum features, such as quantum entanglement, in collider experiments.

\section*{Acknowledgments}
We thank Frank Krauss for his previous collaboration and valuable discussions. We are also grateful to Ansgar Denner and Stefan Dittmaier  for clarifications regarding \texttt{Prophecy4f}. The work of DG, AK, and AN is supported in part by US Department of Energy Grant Number DE-SC 0016013. Some computing for this project was performed at the High Performance Computing Center at Oklahoma State University, supported in part through the National Science Foundation grant OAC-1531128. 
\appendix

\section{Density Matrix Projection and Entanglement}
\label{app:proj}

To assess the validity of neglecting the small negative eigenvalues in the reconstructed density matrix for the $h \to e^+ \nu_e \mu^- \bar{\nu}_\mu$ decay at NLO EW, we project the density matrix $\rho$ onto the nearest positive semi-definite Hermitian matrix with unit trace, denoted $\rho^{\mathrm{proj}}$. The distance between the original and projected matrices is quantified using the Frobenius norm~\cite{10.5555/248979}
\begin{equation}
\left| \rho^{\rm{proj}} - \rho \right|_{F} \equiv \sqrt{\sum\limits_{i,j}\left| \rho^{\rm{proj}}_{ij} - \rho_{ij} \right|^2}.
\end{equation}
We employ the Python package QuTiP~\cite{Johansson:2012qtx} to obtain the projected density matrix that minimizes the  Frobenius norm.

\begin{figure}[!tb]
    \centering
    \includegraphics[width=0.45\textwidth]{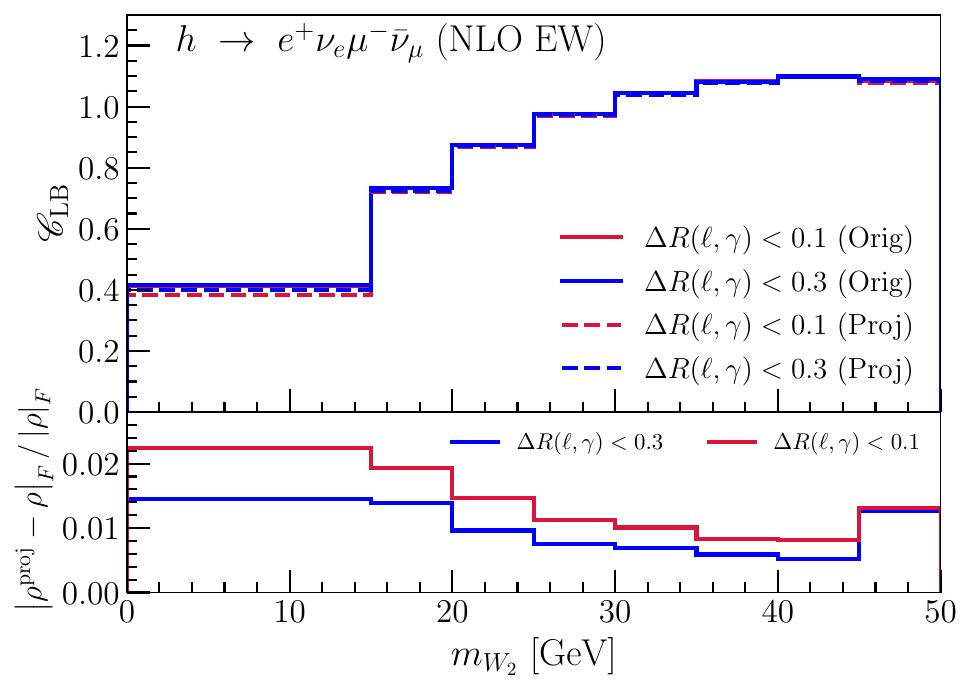}
     \includegraphics[width=0.45\textwidth]{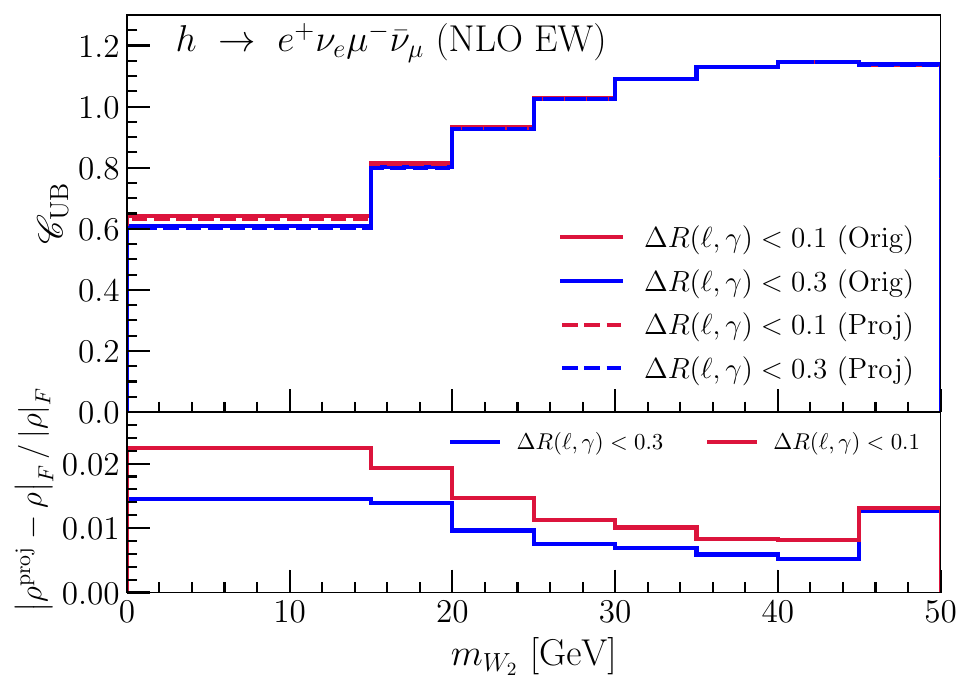}
     \includegraphics[width=0.45\textwidth]{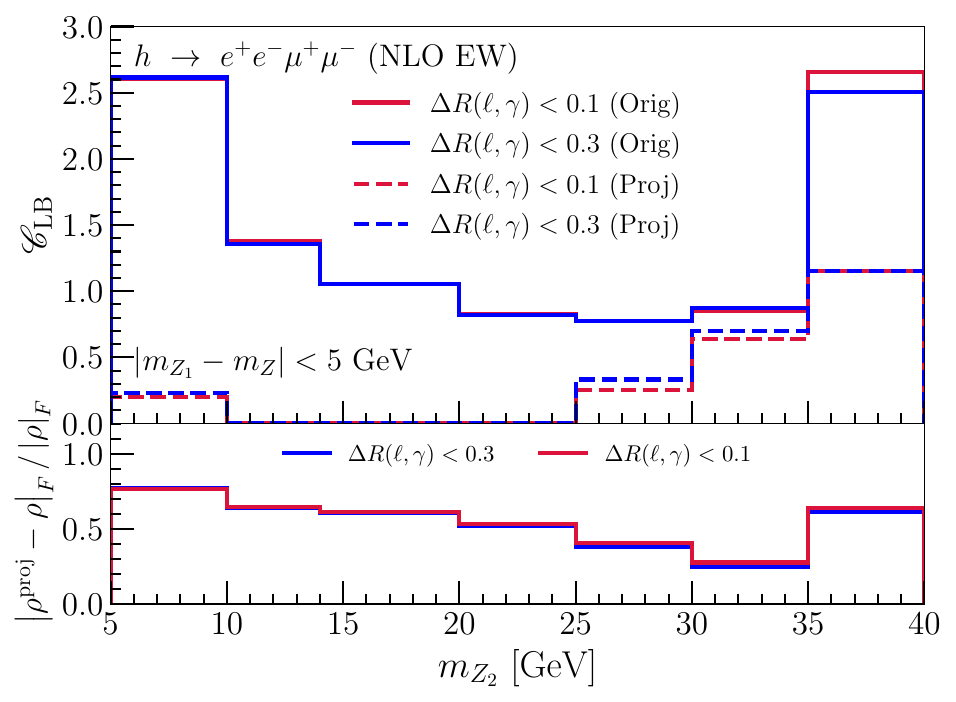}
     \includegraphics[width=0.45\textwidth]{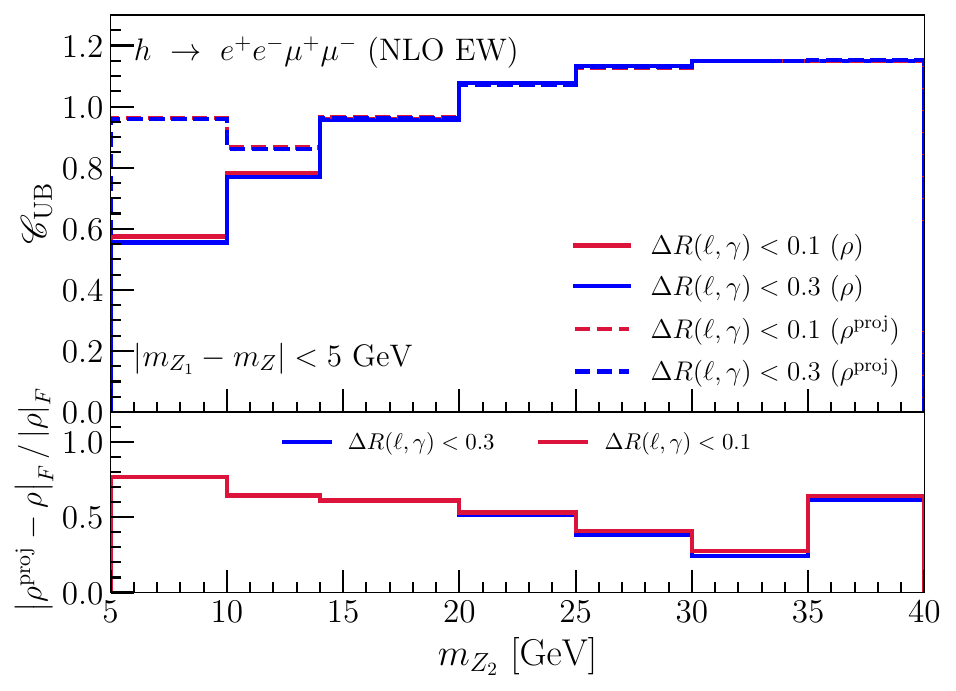}
    \caption{Lower $\mathscr{C}_{\mathrm{LB}}$ (left panel) and upper $\mathscr{C}_{\mathrm{UB}}$ (right panel) bounds of the concurrence for $h\to e^+ \nu_e \mu^- \bar{\nu}_\mu$ (top) and $h\to e^+e^-\mu^+\mu^-$ (bottom) as a function of the off-shell vector boson mass, $m_{V_2}$, with $V=W,Z$. The results are presented at NLO EW with recombination radii $\Delta R(\ell,\gamma)<0.1$ (red) and $\Delta R(\ell,\gamma)<0.3$ (blue). Concurrence bounds computed from the original density matrix $\rho$ are shown as solid lines, while those from the projected matrix $\rho^{\rm{proj}}$ appear as dashed lines. We also present the Frobenius norm between the two matrices, $\left| \rho^{\rm{proj}} - \rho \right|_F$, in the bottom panels of each figure. The results from the original density matrix for $h\to e^+e^-\mu^+\mu^-$ are taken from Ref.~\cite{Goncalves:2025qem}.}
    \label{fig:CboundsNLO_proj}
\end{figure}

In~\autoref{fig:CboundsNLO_proj} (top panel), we compare the concurrence bounds obtained using the original density matrix $\rho$ and the projected matrix $\rho^{\rm{proj}}$. The lower panel of each plot shows the relative distance between $\rho$ and $\rho^{\rm{proj}}$, defined as $\left| \rho^{\rm{proj}} - \rho \right|_F/\left| \rho\right|_F$.
%
We find that the concurrence bounds are nearly identical in both cases. For instance, the difference in $\mathscr{C}_{\rm{LB}}$ is only  $\sim 0.2\%$ 
in the bin $35~\mathrm{GeV} < m_{W_2} < 45~\mathrm{GeV}$, where the magnitude of the smallest negative eigenvalue is greatly suppressed to $\sim -0.005$. A more noticeable difference of about $4\%$ $(7\%)$ in the $\Delta R(\ell,\gamma)<0.3$ (0.1) scenario arises in the first bin, where the negative eigenvalue is comparatively more sizable. To be conservative, we suppress this low $m_{W_2}$ region in~\autoref{fig:CboundsNLO}.

For completeness, we also apply this projection method to $h\to e^+e^-\mu^+\mu^-$ channel, where the reconstructed density matrix exhibits significantly larger negative eigenvalues. The results for the original density matrix for the $h\to e^+e^-\mu^+\mu^-$ channel are taken from Ref.~\cite{Goncalves:2025qem}. In~\autoref{fig:CboundsNLO_proj} (bottom panel), we show the corresponding concurrence bounds and relative Frobenius norms. The sizable differences between the original and projected results for $h\to e^+e^-\mu^+\mu^-$ further underscore that the two-qutrit interpretation of this system is challenged by higher-order corrections.

\section{Dependence on the recombination radius $\Delta R(\ell,\gamma)$}
\label{app:deltaR}

To investigate the impact of real emission on the angular coefficients, we compute these coefficients at NLO EW for various recombination radii, $\Delta R(\ell,\gamma)$. The results are presented in~\autoref{tab:indcoeffsLONLOdR}. In general, the already mild higher-order effects are further suppressed for larger radii. This suppression arises because a larger recombination radius effectively clusters the photon with the corresponding lepton, reducing the distortion of the lepton kinematics induced by real photon emissions. 

\begin{table}[!hb]
\resizebox{\textwidth}{!}{
    \centering
   \begin{tabular}{|c|c|c|c|c|c|}
   \hline 
   \multirow{2}{*}{Coefficient} & \multirow{2}{*}{LO} & \multicolumn{4}{c|}{NLO EW} \\ \cline{3-6}
         &  & $\Delta R(\ell,\gamma)<0.1$ & $\Delta R(\ell,\gamma)<0.3$ & $\Delta R(\ell,\gamma)<1.0$ & $\Delta R(\ell,\gamma)<1.5$ \\ \hline
        $A_{2,0}^{1}$ & $-0.5218(6)$& $-0.5066(6)$& $-0.5095(6)$& $-0.5137(6)$& $-0.5151(6)$ \\
        $A_{2,0}^{2}$ & $-0.5225(6)$& $-0.5068(6)$& $-0.5097(6)$& $-0.5139(6)$& $-0.5152(6)$\\
        $C_{1,0,1,0}$ & $-0.6251(3)$& $-0.6339(3)$& $-0.6327(3)$& $-0.6307(3)$& $-0.6300(3)$\\
        $C_{1,1,1,-1}$ & $ 0.9662(2)$& $0.9674(2)$& $0.9677(2)$& $0.9675(2)$& $0.9672(2)$\\
        $C_{2,1,2,-1}$ & $-0.953(1)$& $-0.961(1)$& $-0.960(1)$& $-0.957(1)$& $-0.956(1)$ \\
        $C_{2,0,2,0}$ & $1.349(2)$& $1.342(2)$& $1.343(2)$& $1.344(2)$& $1.344(2)$ \\
        $C_{2,2,2,-2}$ & $0.599(2)$& $0.604(2)$& $0.603(2)$& $0.599(2)$& $0.596(2)$ \\
        $A_{1,0}^{1}$ & $0$& $-0.0068(3)$& $-0.0047(3)$& $-0.0024(3)$& $-0.0019(3)$ \\
        $A_{1,0}^{2}$ & $0$& $-0.0061(3)$& $-0.0040(3)$& $-0.0018(3)$& $-0.0013(3)$\\
        $C_{1,-1,2,1}$ &  $0$& $-0.0094(7)$& $-0.0059(7)$& $-0.0017(7)$& $0$\\
        $C_{1,0,2,0}$ & $0$& $0.0128(8)$& $ 0.0085(8)$& $0.0031(8)$& $0.0017(8)$ \\
        $C_{2,-1,1,1}$ &  $0$& $-0.0106(7)$& $-0.0072(7)$& $-0.0029(7)$& $-0.0017(7)$\\
        $C_{2,0,1,0}$ &  $0$ & $0.0117(8)$& $0.0072(8)$& $0.0018(8)$& $0$\\
        $C_{1,-1,1,0}$ & $0$ & $0.0010(3)$& $0.0010(2)$& $0.0007(3)$& $0$\\
        $C_{2,-2,2,1}$ &  $0$& $0.005(1)$& $0.005(1)$& $0.004(1)$& $0.003(1)$\\
        \hline 
    \end{tabular}
    }
    \caption{The non-vanishing coefficients for the $h\to  e^+ \nu_e \mu^- \bar{\nu}_{\mu}$ decay at LO and NLO EW for different recombination radius, $\Delta R(\ell,\gamma)$. The invariant mass conditions $|m_{W_1}-m_W|<5$~GeV and $m_{W_2}>10$~GeV are imposed. All coefficients have vanishing imaginary part within uncertainties.}
    \label{tab:indcoeffsLONLOdR}
\end{table}

\end{sloppypar}
\bibliographystyle{utphys}

\bibliography{reference}
\end{document}